# SEISMIC SITE EFFECTS IN A DEEP ALLUVIAL BASIN : NUMERICAL ANALYSIS BY THE BOUNDARY ELEMENT METHOD

J.F.Semblat[1], A.M. Duval[2], P. Dangla[3]

**Abstract :** The main purpose of the paper is the numerical analysis of seismic site effects in Caracas (Venezuela). The analysis is performed considering the Boundary Element Method in the frequency domain. A numerical model including a part of the local topography is considered, it involves a deep alluvial deposit on an elastic bedrock. The amplification of seismic motion (SH-waves, weak motion) is analyzed in terms of level, occuring frequency and location. In this specific site of Caracas, the amplification factor is found to reach a maximum value of 25. Site effects occur in the thickest part of the basin for low frequencies (below 1.0 Hz) and in two intermediate thinner areas for frequencies above 1.0 Hz. The influence of both incidence and shear wave velocities is also investigated. A comparison with microtremor recordings is presented afterwards. The results of both numerical and experimental approaches are in good agreement in terms of fundamental frequencies in the deepest part of the basin. The boundary element method appears to be a reliable and efficient approach for the analysis of seismic site effects.

## 1. SOME DIFFERENT ASPECTS OF EARTHQUAKE GEOTECHNICAL ENGINEERING

Different types of analysis can be performed to investigate earthquake engineering problems. Depending on the purpose of the analysis, one can stress on the only structural aspect of the problem, the geotechnical point of view or, in between, various types of interactions (soil-structure, site-city...). Forgetting the far geological scale including seismic source excitations, there are different scales for the analysis of earthquake geotechnical engineering problems :

- *local geology* : the local topography and sub-surface soil layers are considered, seismic wave propagation in a large (or semi-infinite) domain is the main point to analyze (Bard 1985, Chávez-García 2000, Semblat, 1999, 2000b, Somerville 2000),
- *site-city interaction* : in highly builded areas, some strong interaction can occur between the structures and the subsurface soil layers at the scale of an entire city (Guéguen 2000),
- *soil-structure interaction* : at the scale of several structures, the influence of the soil is mainly involved through energy radiation process, limit conditions, liquefaction... (Anandarajah 1995, Stokoe 1999)

In the following, we will only consider seismic wave propagation and amplification at the local geological scale.

## 2. SITE EFFECTS : EXPERIMENTAL INVESTIGATIONS IN CARACAS

### 2.1 Amplification of seismic motion

The analysis of site effects is very important since the amplification of seismic waves in some specific sites is sometimes strong (Bard, 1985). Reflections and scattering of seismic waves near the surface strengthen the effects of earthquakes (Moeen-Vaziri 1988, Pedersen 1995). The main goal is to avoid, in the design of structures, buildings involving resonant features close to that of the soil surface

---
[1] Res.Eng., La boratoire Central des Ponts et Chaussées (LCPC), Eng. Modeling Dept, 58 bd Lefebvre, 75732 PARIS Cedex 15, France, E-mail : semblat@lcpc.fr
[2] Res.Eng., C.E.T.E Méditerranée, Nice, France
[3] Res.Eng., La boratoire des Matériaux et Structures de Génie Civil (LCPC/CNRS), Marne-la-Vallée, France

layers (Duval 1998, Seed 1970). If the seismic motion amplification is supposed to be strong, the local seismic response of soils must be analyzed to precisely determine the characteristics of the reference earthquake used for the design of structures. Seismic site effects can be investigated using experimental techniques (Duval 1998, Kokusho 1999, Theodulidis 1995) or numerical methods (Beskos 1997, Faccioli 1996, Modaressi 1992, Paolucci 1999, Semblat 1999, 2000a, Wolf 1997). This article mainly investigates the amplification of seismic waves (site effects) through numerical models (boundary element method). It also gives some results (fundamental frequencies) derived from microtremor measurements.

## 2.2 Seismic measurements in Caracas

Caracas is located in northern Venezuela, 16 km from the sea coast. Damages caused in the town by the 1967 earthquake were surprisingly important despite a maximum ground acceleration of around 0.06 to 0.08 g (eastern part). These damages were obviously linked to seismic site effects, because Caracas is built on a wide deep alluvial basin. Many important investigations were made by Seed, Idriss and Dezfulian just after this earthquake (Seed 1970, Duval 1998). Damages due to the 1967 earthquake were concentrated in Palos Grandes for high buildings (14 storeys) and in north-western part of Caracas (less important deposit) for 1 or 2 storeys buildings. There is a link between sediment depth and the height of most damaged buildings.

In 1995, 184 microtremor measurements were performed over the city (Duval, 1998). From these measurements, the deepest part of the deposit is found to give a large H/V ratio, up to a maximum of 6, around 0.6 Hz. This value coincide with the characteristic frequency of damaged buildings in this area. Larger values (around 10) are reached in the southern (thinner) part of the deposit (as previously shown by Seed et al., 1970). Nevertheless, as shown in some other cases (Semblat 1999), H/V ratios derived from microtremor recordings often underestimate the actual amplification factor whereas the fundamental frequencies of alluvial basins are determined with a good accuracy. In this paper, we will then estimate the seismic amplification factor by a numerical analysis and compare afterwards with experimental results (microtremors H/V ratios (Duval 1998)) in terms of fundamental frequency.

## 3. BOUNDARY ELEMENT MODEL

## 3.1 General presentation

To analyze the seismic response of the site, a numerical model based on the boundary element method is considered. The main advantage of this method is to avoid artificial truncation of the domain in the case of infinite medium (Bonnet 1999). For dynamic problems, this truncation could lead to artificial wave reflections giving a numerical error in the solution. Some particular finite element formulations (non-reflecting boundaries, infinite elements) can avoid such problems (Modaressi 1992, Wolf 1997). Nevertheless, finite element or finite difference methods involve some other drawbacks as numerical wave dispersion (Semblat, 2000a) which have to be carefully considered.

The boundary element method can be divided into two main stages (Dangla 1988, Bonnet 1999) :
− solution of the boundary integral equation giving displacements and stresses along the border of the domain,
− a posteriori computation for all points inside the domain using an integral representation formula.

One considers an elastic, homogeneous and isotropic solid of volume $\Omega$ and external surface $\partial\Omega$. The problem is supposed to have an harmonic dependence on time (circular frequency $\omega$). The displacement field is then written $u(x,t)=u(x).e^{-i\omega t}$. The displacement magnitude $u(x)$ is solution of the following equation :

$$(\lambda + 2\mu)\,grad\,(div\ u(x)) - \mu\ rot(rot\ u(x)) + \rho\ f(x) + \rho\omega^2 u(x) = 0 \tag{1}$$

where $u$ is the displacement field, $\lambda$ and $\mu$ are the Lamé constants and $f$ a body force.

## 3.2 Integral formulation

For steady state solutions of harmonic problems, the reciprocity theorem gives an integral equation between two displacement fields $u$ and $u'$, solutions of equation (1), corresponding to mass forces $f$ and $f'$ (resp.). This relationship can be written as follows:

$$\int_{\partial\Omega} t^{(n)}(x)u'(x)ds(x) + \int_{\Omega} \rho f(x)u'(x)dv(x) = \int_{\partial\Omega} t'^{(n)}(x)u(x)ds(x) + \int_{\Omega} \rho f'(x)u(x)dv(x) \qquad (2)$$

The integral formulation arises from the application of the reciprocity theorem between the unknown displacement field and the fundamental solutions of a reference problem called Green's functions. The reference problem generally corresponds to the case of an infinite space (or semi-infinite for SH-waves) in which a body force concentrated at point $y$ acts along direction $e$: $\rho f'(x) = \delta(x-y)e$.

Green's functions are denoted $U_{ij}^{\omega}(x, y)$ and represent the complex displacement along direction $j$ at point $x$ due to a concentrated unit force at point $y$ along direction $i$. The related stress field, considering the normal $n(x)$, is denoted $T_{ij}^{(n)\omega}(x, y)$. The application of the reciprocity theorem between $u(x)$ and the Green function $U_{ij}^{\omega}(x, y)$ then leads to the following integral representation (Bonnet, 1999, Dangla, 1988):

$$I(y)u_i(y) = \int_{\partial\Omega} \left( U_{ij}^{\omega}(x,y)t_j(x) - T_{ij}^{(n)\omega}(x,y)u_j(x) \right) ds(x) \qquad (3)$$

$I(y)$ is 1 if $y \in \Omega$ and 0 otherwise, $u_j$ is the displacement along $j$, $t_j$ is the $j$ component of the traction vector applied on surface $\partial\Omega$.

This equation allows for the estimation of the displacement field $u$ in every point inside the volume $\Omega$. One should have nevertheless determine the displacement field and traction vector in all points of the surface $\partial\Omega$. It is then necessary to solve the surface integral equation (Bonnet, 1999, Dangla, 1988) which can be written as follows:

$$\theta_{ij}(y)u_j(y) + \int_{\partial\Omega} T_{ij}^{(n)\omega}(x,y)u_j(x)ds(x) - \int_{\partial\Omega} U_{ij}^{\omega}(x,y)t_j(x)ds(x) = 0 \qquad (4)$$

where $\theta_{ij}(y) = \delta_{ij}/2$ if the contour is regular with $y$. If the contour is not regular with $y$ (actual case when the problem is discretized) $\theta_{ij}(y)$ is written in a different form (Bonnet, 1999, Dangla, 1988).

Numerical solution of equation (3) can be estimated by collocation method, or thanks to a variational formulation (Bonnet, 1999). Equations (3) and (4) are written for a finite volume. When the domain $\Omega$ is infinite, it is necessary to include some radiation conditions (called Sommerfeld's conditions) leading to a divergent diffracted wave (Dangla, 1988). The diffracted displacement field is defined by the relation $u^d = u - u^{inc}$ where $u^{inc}$ represents the incident wave coming from infinity for which the Sommerfeld's conditions can obviously not be applied.

In this article, the dynamic problem is analyzed in two dimensions (plane strain). Two-dimensional Green's functions of infinite medium can be expressed thanks to Hankel's functions. The solution of equation (3) is classically obtained by finite boundary elements discretization and then by collocation, that is application of the integral equation at each node of the mesh (Bonnet, 1999).

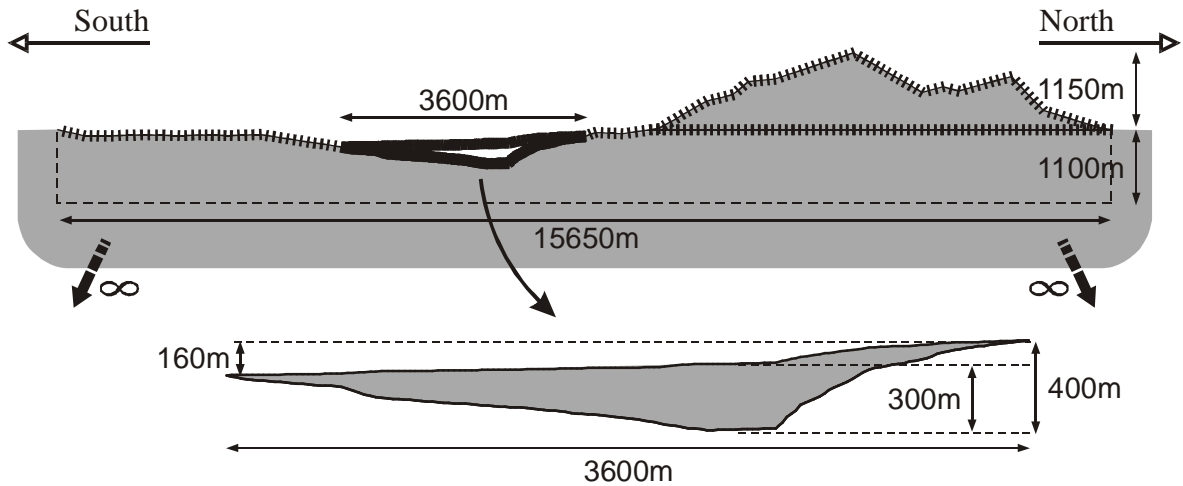

**FIG. 1. Boundary element mesh for a plane SH wave
(North-South profile of Palos Grandes, Caracas).**

*3.3 Modelling of the geological profile*

For the numerical simulation in Caracas, a North-South geological section corresponding to Palos Grandes is considered (Figure 1). The alluvial layer is supposed to be homogeneous. The mechanical characteristics of the deposit and of the bedrock (density, shear modulus and corresponding velocity) are the following :
− alluvial layer : $\rho_1$=2000 kg/m$^3$, $\mu_1$=405 MPa, giving $C_1$=450 m/s,
− elastic bedrock : $\rho_2$=2300 kg/m$^3$, $\mu_2$=14375 MPa, giving $C_2$=2500 m/s.

The thickest part of the alluvial deposit has a maximum depth of around 300 m at a distance of 2150 to 2500 m of the southern edge. The basin depth decreases fastly on the northern part and much slowlier on the southern part. The loading is an harmonic plane SH-wave with vertical incidence (first) and various incidences (afterwards). There is then only one displacement component orthogonal to the model plane.

In Figure 1, the boundary element mesh is depicted for the solution of the integral equation (wave propagation in a semi-infinite medium). Its total width is around 15 km for a basin of approximately 3600 m. The solution is computed a posteriori at some other points inside the domain. This model uses Green's functions of an infinite domain for the alluvial layer and the mount. It involves Green's functions of a semi-infinite domain (easy to determine for SH-waves) to precisely model the bedrock as a subdomain of an infinite half-space. Elastodynamics equations are solved in the frequency domain and the computation is made using the FEM/BEM code CESAR-LCPC (Humbert 1989).

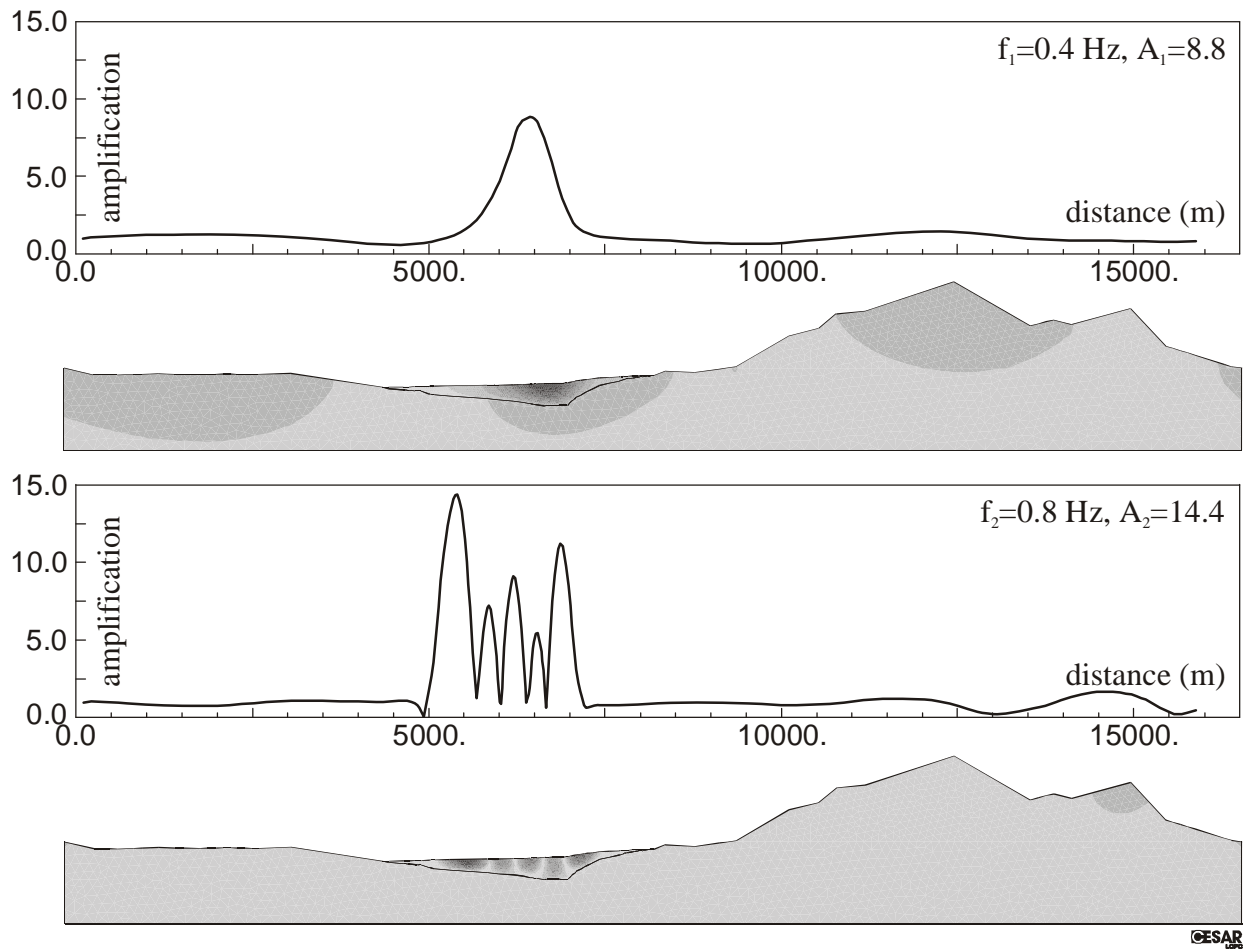

**FIG. 2. Estimation of the amplification factor along the whole profile for a vertical SH-wave (variable grey scale).**

## 4. ESTIMATION OF AMPLIFICATION

### 4.1 Amplification factor for a vertical incidence

The amplification factor is estimated from the computed displacements. Considering the whole geological profile (Figure 2), the maximum amplification obviously occurs in the deep alluvial basin. The perturbation of the wavefield due to the local topography (North) is negligible when compared with site effects in the basin itself mainly due to the impedance ratio between the basin and the bedrock. The influence of the basin geometry on the amplification level could also be very strong (Semblat 1999).

The local amplification of seismic motion is much stronger in the alluvial deposit (Figure 3). Site effects firstly occur in the thickest part of the basin for low frequencies. For higher frequencies, different areas of high amplification appear on both sides of the deposit. Above 1.2 Hz, high amplification areas are not only located on the surface but also inside the deposit itself. There is a strong focusing effect leading to strong amplification areas within the basin itself (Somerville 2000). The variations of the amplification factor are very large since it ranges from 6.5 to 20.

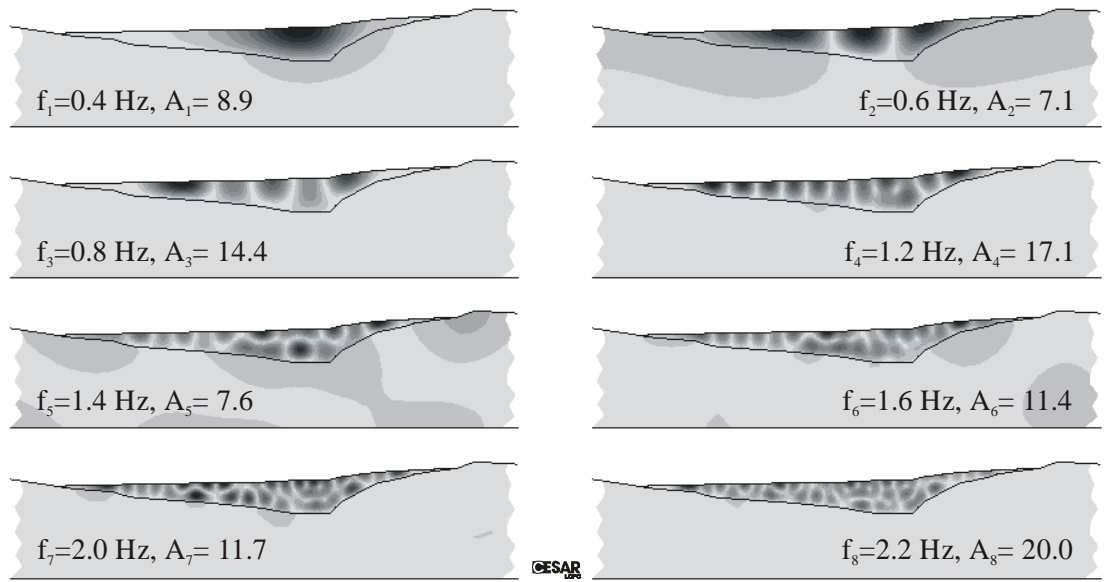

**FIG. 3. Estimation of the amplification factor in the alluvial basin
for a vertical SH-wave at different frequencies (variable scale).**

*4.2 Analysis of location and frequency*

To fully characterize site effects, it is necessary to investigate amplification level versus both location and frequency. As shown in Figure 4, amplification appears first in the thickest part of the deposit for low frequencies. For higher frequencies, the maximum amplification area is located further and further from this deep zone on southern and northern parts of the basin. A strong amplification is found around 0.8 Hz with a factor above 25 at approximately 2500 m from the southern edge of the basin. These results are in agreement with experimental results found by Duval (1998). The numerical results also show two other high amplifications between 1 and 1.5 Hz (at 1500 m) and above 2 Hz (at 1300 and 2200 m).

*4.3 Maximum overall amplification vs frequency*

From these numerical results, one can display the maximum overall amplification factor, that is the maximum amplification, at each frequency, all along the surface (no reference is then made to the location). As indicated in Figure 5, the maximum overall amplification factor fastly changes with frequency. Many important amplifications between 10 and 30 are observed for various frequencies. The three major site effects are detected around 0.6 Hz, 0.8 and 0.9 Hz with respective amplification factors of 20, 27 and 26. Some other amplifications are found for higher frequencies but we will analyze the corresponding level for both damped and undamped cases in the next section.

The frequency dependence of amplification is rather different from that of other geological profile. Comparatively, in the case of Nice (French Riviera) for a very shallow basin, the amplification factor was found very small for low frequencies (below 1 Hz), became progressively larger and remained large for higher frequencies (Semblat 1999). This is probably due the shape of both geological profiles : very flat and regular in the case of Nice (Semblat 2000b), deeper and irregular for Caracas. These differences are obvious when considering the focusing effect in the case of Caracas and the amplification process within the deep basin itself (Fig. 3).

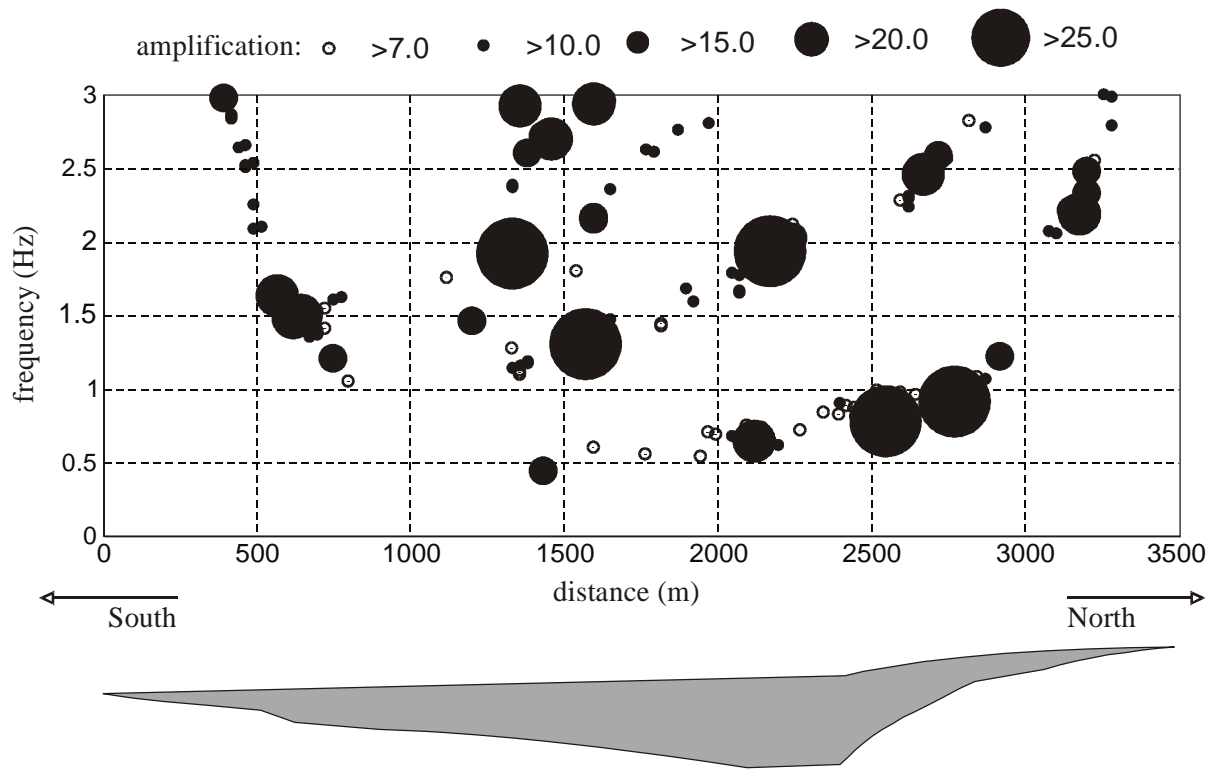

**FIG. 4. Amplification along the basin surface versus distance and frequency.**

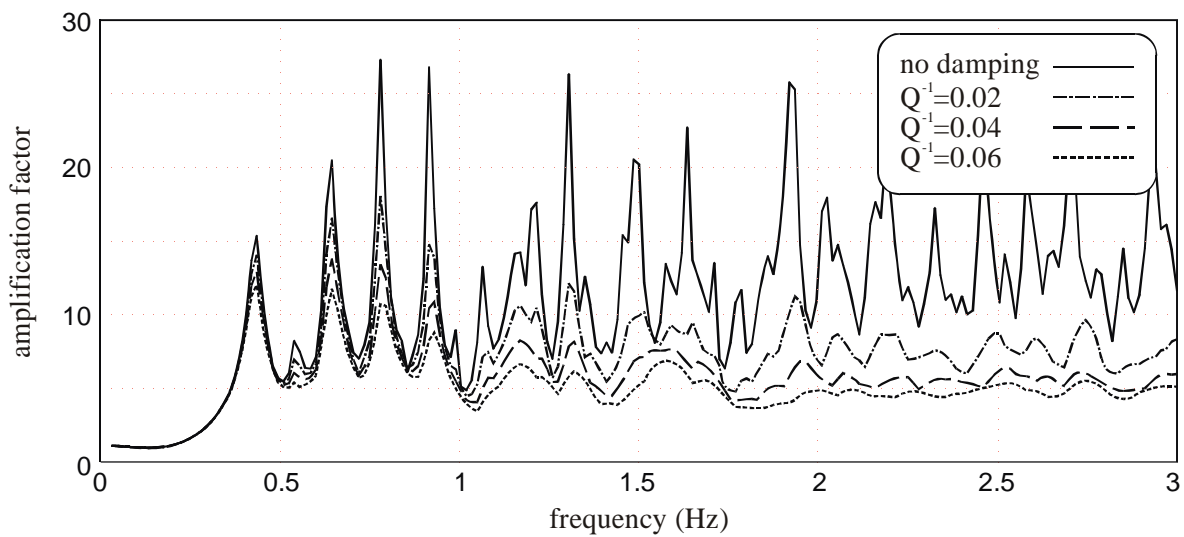

**FIG. 5. Maximum overall amplification at the basin surface
versus frequency for different damping values.**

## 5. AMPLIFICATION IN A DAMPED ALLUVIAL BASIN

The basin is now assumed as a damped linear elastic medium. Its mechanical behaviour corresponds to a Zener (standard solid) rheological model. This model is defined in the following references (Bourbié 1987, Shames 1992, Semblat 1997, 98). The peak of the attenuation-frequency curve (Bourbié 1987) is defined at 2 Hz and different values of maximum attenuation are chosen : $Q^{-1}$=0.02, $Q^{-1}$=0.04 and $Q^{-1}$=0.06 where $Q$ is the quality factor. For the boundary integral formulation considered here, the computations are performed in the frequency domain. The damping properties are then involved through the complex modulus of the Zener viscoelastic model (Bourbié 1987, Semblat 1998).

Figure 5 gives the maximum overall amplification factor in the elastic case and in the three damped cases defined previously. These curves show that the three main amplifications appearing below 1 Hz in the undamped case are also found in the damped cases. They appear at the same frequencies with smaller amplification levels but still always above a factor of 10. For higher frequencies, the amplification factor is much lower in the three damped cases leading to more realistic amplification-frequency curves. For these frequencies, the influence of the large depth of the (damped) alluvial basin on the maximum amplification peaks is then strong.

## 6. INFLUENCE OF INCIDENCE

As the bedrock is described using Green's functions of an infinite half-space, seismic amplification can be determined, in the case of plane SH-waves, for every values of incidence. To analyze the effect of incidence, results are firstly given in a specific point versus frequency.

Results of Figure 6 give the values of the amplification factor in all points of the surface of the deposit for two specific frequency values (1.0 Hz left, 1.2 Hz right). Three different incidence values are considered : 60, 90 and 120° (see Figure 6). It can be noticed that the amplification factor at 1.0 Hz is much higher for 60° and 120° incidences (up to 20) than for a normal incidence (around 7.0). At 1.2 Hz, the amplification is stronger for normal incidence (up to 17) and is maximum near southern and northern edges of the basin (600 and 3000 m from southern edge). At the same frequency, both other incidences give lower amplification (from 7 to 10). The location of the maximum amplification is also different and is found in the deepest area of the deposit (2200 m from southern edge). At 1.2 Hz, both results (for symmetrical incidence angles) have significant discrepancy. There is however no significant difference at 1.0 Hz. The influence of incidence is then observed on magnitude of amplification, occuring frequency and location of the corresponding area.

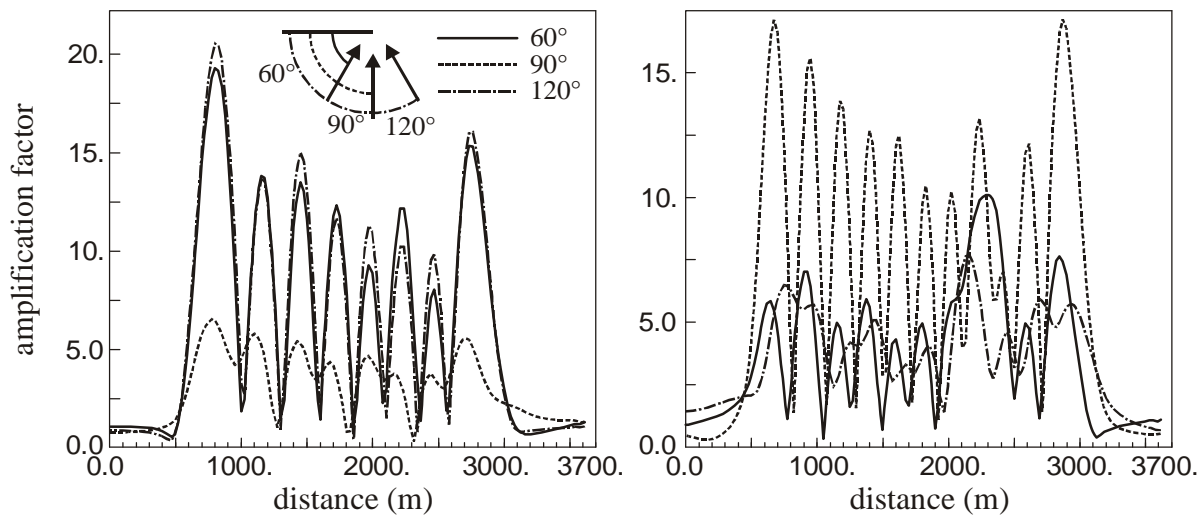

**FIG. 6. Dependence of the amplification factor on incidence for two different frequencies (left: 1.0Hz, right: 1.2Hz).**

## 7. INFLUENCE OF THE MECHANICAL PROPERTIES

In these numerical computations, it is also interesting to investigate the influence of the mechanical parameters on amplification levels. Figure 7 displays three amplification-frequency curves (in logscale) corresponding to three different values of shear modulus for the alluvial basin. The previous value $\mu_1$=405 MPa corresponds to case (a), the values of shear moduli for cases (b) and (c) are respectively : $\mu_b=2\mu_1/3$ and $\mu_c=1.5\mu_1$. The amplification levels reached in cases (b) and (c) are significantly larger and smaller (resp.) than in case (a). The main difference is nevertheless the increasing part of the curves showing a much faster increase of the amplification factor in case (b) than in case (a) and in case (a) than in case (c). The sensitivity of the amplification factor to the mechanical properties is very important for this feature of the level-frequency curves. Concerning the maximum values of the amplification factor, one needs some experimental investigations from real earthquake measurements (Standard Spectral Ratios) to find out the most accurate case.

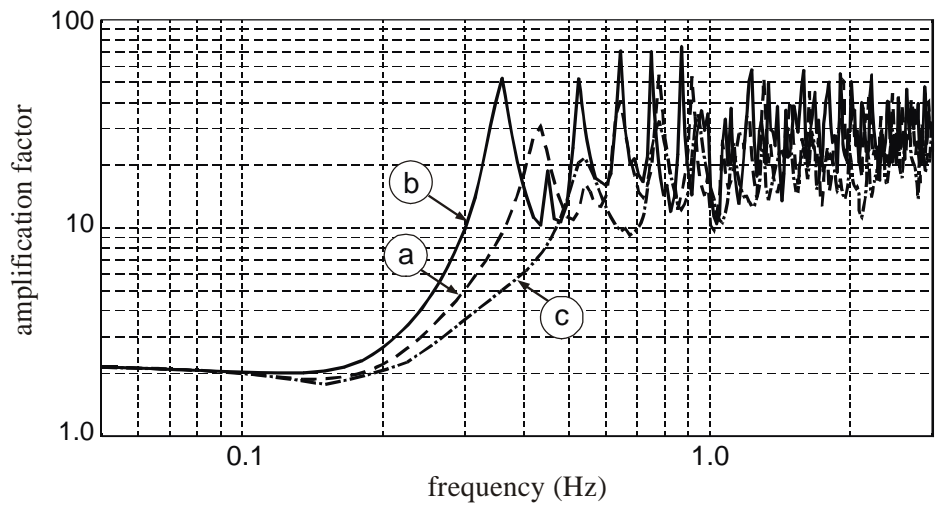

**FIG. 7. Maximum overall amplification along the surface for different shear moduli:
(a) $\mu$=405MPa, (b) $\mu$=270Mpa and (c) $\mu$=608MPa.**

## 8. COMPARISON WITH EXPERIMENTAL MEASUREMENTS

The microtremors recordings are known to give accurate values of fundamental frequencies but the link between H/V ratios and actual amplification factors is not always clear. In the case of Nice in France (Semblat 1999), H/V ratios were found to be much smaller than Standard Spectral Ratios from real earthquake measurements. On the contrary, in this case, the agreement between numerical amplification factors and SSR was found to be very good (Semblat 2000b). Hopelessly, in the case of Caracas, we do not have some estimations of the SSR from earthquake measurements (Duval 1998). The comparison between BEM results and H/V ratios from microtremor recordings is then proposed here in terms of fundamental frequencies for antiplane (E-W) motion corresponding to the SH-wave excitation in the BEM model.

Figure 8 gives points (distance-frequency) for which the amplification factor is above a level of 7.0. There is a rather good agreement with numerical results for main amplification areas in the center of the basin with very close values for both distance and frequency. For amplifications at higher frequencies, the comparison is rather good. However, some higher frequency amplifications are found from the numerical results which can not be recovered by microtremors H/V ratios. This discrepancy can be explained by the fact that this experimental technique generally gives the so-called fundamental frequency with only one single peak on H/V spectral ratio curves. It is therefore not possible with this experimental technique to find several amplifications at the same location but for different frequencies. There is not such a limitation with the numerical method considered herein.

Experimental values in terms of amplitudes are much lower than numerical ones (Duval 1998) : mainly below 6 or 8 compared with 20 or 25 for boundary element method computations. In some previous work, this strong discrepancy was found to be due to the limitations of the microtremor technique. Numerical BEM models gave maximum amplification factors very close to experimental Standard Spectral Ratios estimated from real earthquake measurements (Semblat 1999).

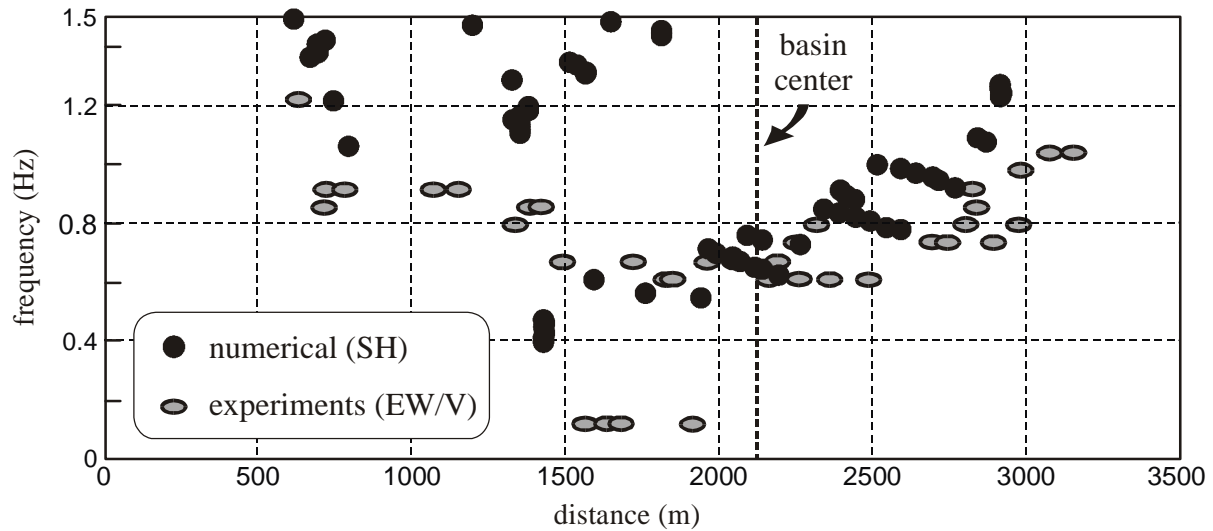

**FIG. 8. Comparison between numerical and experimental results
in terms of fundamental frequencies (East-West component).**

## 9. CONCLUSION

In situ microtremor measurements performed in Caracas by the CETE-Méditerranée (Duval 1998) show that the seismic motion is strongly amplified above 0.6 Hz. The corresponding area is located above the deepest part of the alluvial deposit.

Numerical simulations (SH-waves) based on the boundary element method are in good agreement with these results. For higher frequencies, the maximum amplification area is found to be on both southern and northern sides of the basin. It is located further away from the center of the deposit when frequency increases. Both methods (experimental, numerical) identically lead to this conclusion.

Numerical computations nevertheless give higher amplification factors (from 20 to 27). As for other cases (Semblat 1999), the geometry of the alluvial basin has a strong influence on the amplification level which is not very well estimated by microtremor H/V ratios.

This article has presented the influence of the mechanical properties of the basin (damping, shear modulus) on the amplification-frequency dependence. Site effects also appear to be sensitive to incidence. It changes the maximum amplification factor, the frequency at which it occurs and the corresponding affected area.

From the numerous numerical results obtained and the good agreement with experimental ones in terms of fundamental frequencies, the boundary element method seems to be efficient and reliable to analyze site effects from a qualitative, as well as quantitative, point of view. Some further experimental investigations considering Standard Spectral Ratios from real earthquake measurements could give some interesting information for a complete comparison with the BEM numerical results with respect to the amplification level.